# Anharmonicity in the High-Temperature *Cmcm* Phase of SnSe: Soft Modes and Three-Phonon Interactions

Jonathan M. Skelton, Lee A. Burton, Stephen C. Parker and Aron Walsh*
*Department of Chemistry, University of Bath, Claverton Down, Bath BA2 7AY, UK*

Chang-Eun Kim and Aloysius Soon
*Department of Materials Science and Engineering, Yonsei University, Seoul 120-749, Korea*

John Buckeridge, Alexey A. Sokol and C. Richard A. Catlow
*University College London, Kathleen Lonsdale Materials Chemistry, Department of Chemistry, 20 Gordon Street, London WC1H 0AJ, United Kingdom*

Atsushi Togo and Isao Tanaka
*Elements Strategy Initiative for Structural Materials, Kyoto University, Kyoto Prefecture 606-8501, Japan*

The layered semiconductor SnSe is one of the highest-performing thermoelectric materials known. We demonstrate, through a first-principles lattice-dynamics study, that the high-temperature *Cmcm* phase is a dynamic average over lower-symmetry minima separated by very small energetic barriers. Compared to the low-temperature *Pnma* phase, the *Cmcm* phase displays a phonon softening and enhanced three-phonon scattering, leading to an anharmonic damping of the low-frequency modes and hence the thermal transport. We develop a renormalisation scheme to quantify the effect of the soft modes on the calculated properties, and confirm that the anharmonicity is an inherent feature of the *Cmcm* phase. These results suggest a design concept for thermal insulators and thermoelectric materials, based on displacive instabilities, and highlight the power of lattice-dynamics calculations for materials characterization.

Thermoelectrics are an important class of functional materials which interconvert heat and electricity,[1] and are a key component in the drive for "green" energy.[2] The figure of merit for thermoelectric performance is $ZT = S^2 \sigma T/(\kappa_L + \kappa_{el})$, where $S$ and $\sigma$ are the Seebeck coefficient and electrical conductivity, and $\kappa_L$ and $\kappa_{el}$ are the lattice and electronic thermal conductivities, respectively. The key to developing high-performance thermoelectric materials is to reduce the thermal conductivity while maintaining a high thermopower $S^2\sigma$. Widespread application requires a $ZT$ above 2 at the target operating temperature.[3]

Historically, the lead chalcogenides set the benchmark for high thermoelectric performance due to their unique combination of favourable electrical properties and strongly-anharmonic lattice dynamics.[4-11] However, bulk SnSe was recently shown to be a very promising high-temperature thermoelectric, with a $ZT$ score surpassing the record set by nanostructured PbTe.[12] This has been ascribed to an ultra-low lattice thermal conductivity, arising from its pseudo-layered structure and strong high-temperature anharmonicity.[13,14] Compared to PbTe, SnSe achieves superior thermoelectric performance without the need for doping or other material modifications,[15] which can be detrimental to electrical properties. It is therefore important to elucidate the microscopic origin of its high performance, in order that the same ideas may be incorporated into design strategies for improving thermoelectrics.

SnSe displays a second-order phase transition from *Pnma* to *Cmcm* phases between 700-800 K.[13,14] Modelling studies on the thermal transport of the low-temperature phase have been carried out,[16] while combined theoretical and computational studies of both phases[14,17] have connected the low lattice thermal conductivity to the anharmonicity arising from the displacive phase transition. However, there has not yet been a detailed comparison of the intrinsic lattice dynamics of the two phases, in part due to the difficulties associated with modelling phonon instabilities.

Previous theoretical studies have shown that first-principles lattice-dynamics calculations can provide deep insight into the material physics underlying thermoelectric performance.[18-25] In this work, we have employed first-principles lattice-dynamics calculations to model both phases of SnSe, with a particular focus on characterising the lattice dynamics and thermal transport in the high-temperature phase, and on developing a practical approach to the theoretical challenge of modelling phonon soft-mode instabilities.

Our lattice-dynamics calculations were performed using the Phonopy[26,27] and Phono3py[28] packages, with the VASP code[29] as the force calculator. We used the PBEsol functional,[30] which we have previously found to give a very good description of the lattice dynamics in a range of semiconductors.[31] Projector augmented-wave (PAW) pseudopotentials[32,33] treating the Sn 4d, 5s and 5p and the Se 4s and 4p electrons as valence were used to model the ion cores. The electronic wavefunctions were expanded in a plane-wave basis with a kinetic-energy cutoff of 500 eV. An 8×4×8 Monkhorst-Pack *k*-point mesh was used to sample the Brillouin zone, correspondingly reduced for the supercell force calculations.[34] The second-order (harmonic) force constants were computed using a carefully-chosen 6×1×6 supercell expansion for both phases and a displacement step of $10^{-2}$ Å. The third-order force constants were computed using 3×1×3 expansions and a $3×10^{-2}$ Å step. Full technical details of our simulations, along with supercell-size convergence tests for the harmonic-phonon calculations, can be found in the supporting information.[41]

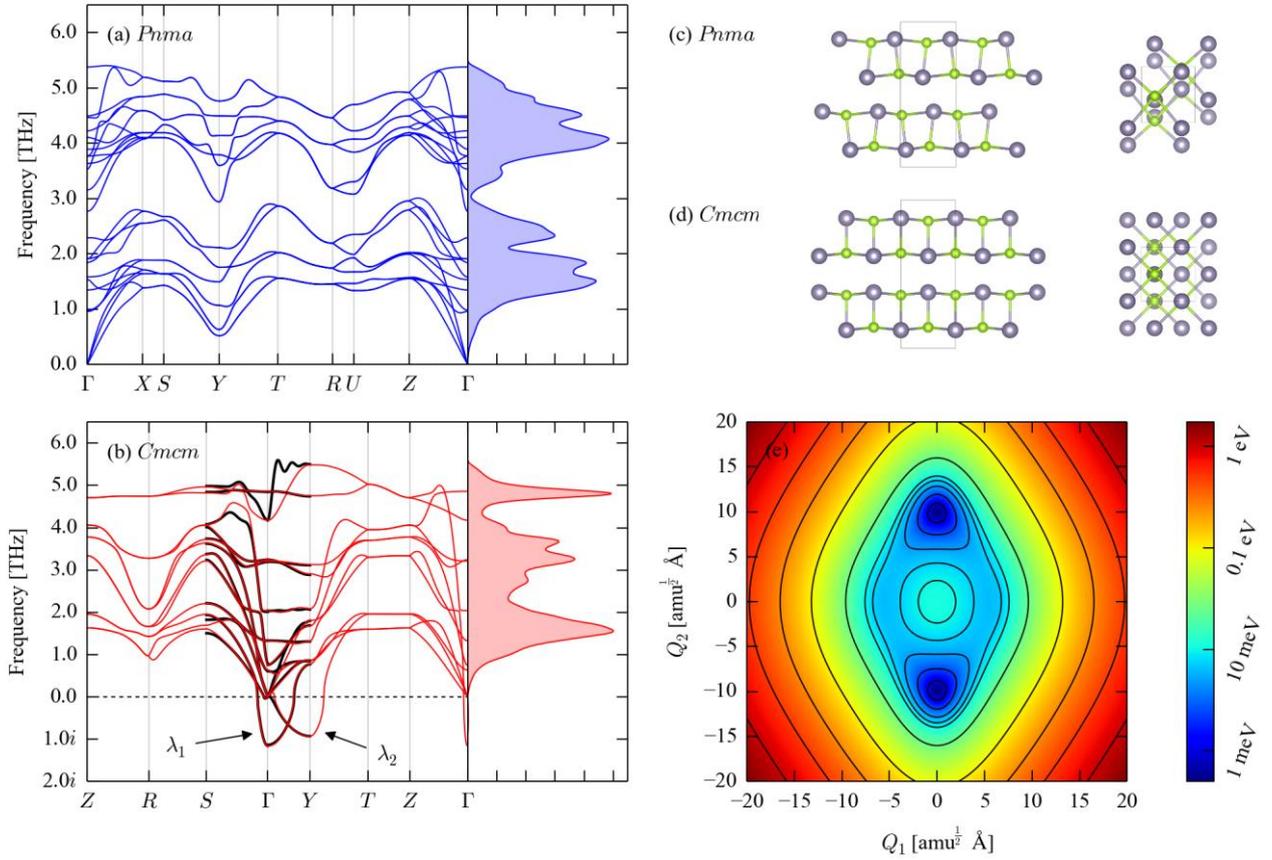

FIG. 1 (Color Online) Lattice dynamics of SnSe. Plots (a) and (b) show the phonon dispersions and densities of states of the optimised equilibrium *Pnma* and *Cmcm* structures illustrated in (c) and (d), respectively. The dispersion of the high-temperature *Cmcm* phase shows imaginary modes at the symmetry points $\Gamma$ and $Y$, marked $\lambda_1$ and $\lambda_2$ respectively. The black lines in the *Cmcm* dispersion are segments computed from supercells chosen to have a larger number of commensurate points along the corresponding reciprocal-space directions (Ref. [41]). The 2D potential-energy surface along the two corresponding normal-mode coordinates $(Q_1/Q_2)$ is shown in plot (e). The left- and right-hand snapshots in (c)/(d) are taken through the *ab* and *ac* planes, respectively, and the Sn and Se atoms in the models are coloured silver and green. The images were prepared using VESTA (Ref. [35]).

The optimised equilibrium structures of the two phases are shown in Figs. 1c and 1d. The calculated lattice parameters of the *Pnma* and *Cmcm* phases are $a = 4.367$, $b = 11.433$ and $c = 4.150$ Å, and $a = 4.217$ Å, $b = 11.525$ and $c = 4.204$ Å, respectively. These values are consistent with experimental characterization (*Pnma*: $a = 4.44$ Å, $b = 11.49$ Å, $c = 4.135$ Å; *Cmcm*: $a$, $c = 4.31$ Å, $b = 11.70$ Å), and with other theoretical studies.[13,16] The low-temperature phase is a distortion of the more symmetric high-temperature structure, leading to a primitive unit cell with twice the volume, and both structures can be thought of as distortions to the rocksalt structure of the Pb analogue PbSe.

Comparing the phonon dispersion and density-of-states (DoS) curves of the two phases (Figs. 1a/1b), the upper-frequency part of the *Cmcm* DoS displays a noticeable red shift with respect to the *Pnma* phase. From a numerical integration,[41] around 15 % of the states in the low-temperature phase lie between 3-4 THz, compared to ~30 % in the high-temperature phase. Moreover, whereas the dispersion of the *Pnma* structure shows real frequencies across the Brillouin zone (Fig. 1a), the dispersion of the high-temperature phase displays prominent imaginary modes along the line between $\Gamma$ and $Y$ (Fig. 1b), with the primary soft modes at the two symmetry points.[14,17]

The imaginary modes make an almost negligible contribution to the overall phonon DoS, suggesting them to be confined to a small reciprocal-space volume. This was confirmed by inspecting the frequencies at the commensurate **q**-points in the various supercells used for convergence testing, as well as in a set of 2×$N$×2 expansions with a high density of commensurate points along the $\Gamma$-$Y$ line.[41]

From an analysis of the phonon eigenvectors,[41] both modes correspond to symmetry-breaking displacive instabilities. The optic mode at $\Gamma$ corresponds to motion of the four (two) Sn and Se atoms in the conventional (primitive) cell in opposing directions parallel to the $c$ axis. The acoustic mode at $Y$ is a similar motion along the $a$ axis, but with the atoms in the two bonded layers moving in opposite directions. From mapping out the 2D potential-energy surface spanned by the two mode eigenvectors (Fig. 1e), it can be seen that the $Y$ mode represents the primary distortion to the low-temperature phase: the minimum along the $\Gamma$-point mode ($\lambda_1$ in Fig. 1b/$Q_1$ in Fig. 1e) is another saddle point on this potential-energy surface, whereas the global energy minimum lies along the $Y$ mode ($\lambda_2/Q_2$). It can also be seen from a comparison of the *Pnma* and *Cmcm* structures in Fig. 1c/d that the symmetry breaking induced by this mode maps between the high-temperature to the low-temperature phase.

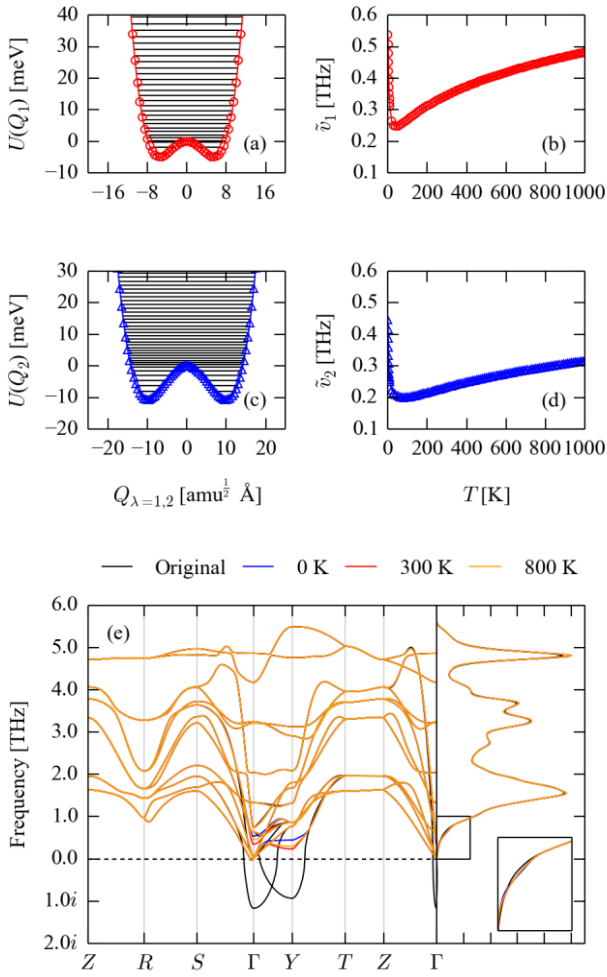

FIG. 2 (Color online) Renormalization of the imaginary harmonic-phonon modes in the equilibrium *Cmcm* SnSe structure. Plots (a) and (c) show the anharmonic double-well potentials along the two modes labelled $\lambda_1$ and $\lambda_2$ in Fig. 1b as a function of the corresponding normal-mode coordinates $Q_{\lambda=1,2}$. The markers show the calculated points, the solid lines are fits to a 20-power polynomial, and the black lines inside the potentials show the eigenvalues obtained by solving a 1D Schrödinger equation. Plots (b) and (d) show the effective renormalized frequencies of the two modes ($\tilde{\nu}_{\lambda=1,2}$) as a function of temperature, calculated as the harmonic frequencies which reproduce the contributions of the modes to the vibrational partition function. Plot (e) compares phonon dispersions and densities of states calculated without renormalization (black) and with the imaginary modes renormalized to the calculated 0 (blue), 300 (red) and 800 K (orange) frequencies.

Each imaginary mode forms an anharmonic double-well potential (Figs. 2a/2c). The potential along the Γ-point mode is steeper, accounting for its more negative harmonic frequency, but, as seen in Fig. 1e, the potential along the *Y* mode possesses a deeper minimum. The respective minima are both shallow, at 5 and 11 meV below the average structure. These findings indicate that the *Cmcm* phase is effectively an average structure, in a manner analogous to the high-temperature phases of some oxides[36] and oxide/halide perovskites (e.g. CsSnI₃).[25,37,38] To confirm that the *Cmcm* phase is unlikely to become stable under lattice expansion (or contraction) at finite temperature, we computed phonon dispersions and DoS curves over a range of expansions and contractions about the athermal equilibrium volume,[41] which indicated that the imaginary modes soften further under lattice expansion, and also persist under moderate applied pressure.

Soft modes present a challenge to lattice-dynamics calculations, as techniques for treating this anharmonicity (e.g. self-consistent phonon theory[39,40]) are often impractically expensive. Whereas in real systems phonon instabilities are localized to regions of infinitesimal volume in reciprocal space (i.e. points, lines or planes), and thus do not contribute to thermodynamic properties such as the free energy,[36] the finite Brillouin-zone sampling and interpolation used in lattice-dynamics calculations "spreads" the imaginary modes over a finite volume.

In order to investigate the effect of the imaginary modes in the high-temperature phase on our calculations, we implemented a simple renormalization scheme to calculate approximate effective harmonic frequencies. The potential energy along each mode as a function of the respective normal-mode coordinate, $Q_{\lambda=1,2}$, is fitted to a 20-power polynomial and 1D Schrödinger equations for the potentials are solved to obtain the eigenvalues inside the anharmonic double wells (Figs. 2a/2c). Effective (real) harmonic frequencies for the imaginary modes are then calculated as a function of temperature to reproduce the contribution of the modes to the anharmonic vibrational partition function (Figs. 2b/2d). These effective frequencies are used to adjust the harmonic force constants, by back transforming the dynamical matrices at the commensurate **q**-points in the supercell expansion after adjusting the frequencies of the imaginary modes. The corrected force-constant matrices are then used as input for further post processing. A detailed overview of the renormalization scheme and its implementation is given in the supporting information.[41]

The Γ and *Y* points in the *Cmcm* primitive cell are commensurate with our chosen 6×1×6 supercell expansion, but the renormalization has no commensurate points along the Γ-*Y* line. The dispersion appears to be well reproduced with this expansion (c.f. Fig. 1b), however, and it was thus employed for renormalization at the symmetry points. The renormalized 0 K frequencies of the imaginary modes at Γ and *Y* are 0.54 and 0.45 THz, respectively. The frequencies initially decrease with temperature, as the system explores higher energy levels within the minima, and reaches a minimum close to where the available thermal energy is comparable to the barrier height (approx. 40 and 80 K, respectively, for the two modes). Above these temperatures, the renormalized frequencies increase monotonically as the system accesses the steeper parts of the potential. The renormalized frequencies at 300 K are 0.35 (Γ) and 0.23 THz (*Y*), and increase to 0.45 (Γ) and 0.30 THz (*Y*) at 800 K (the temperature above which the *Cmcm* phase is observed crystallographically[13,14]).

Fig. 2e shows the impact of renormalizing the imaginary modes at 0, 300 and 1000 K on the phonon dispersion and DoS. The effect on the orthogonal phonon branches is minimal, and the correction is strongly localized to the lines along which the imaginary modes occur. The latter is primarily a result of the large supercell used for calculating the force constants, which we consider a prerequisite for this renormalization scheme. The effect of the renormalization on the phonon DoS is minimal, with the most notable differences occurring up to ~1 THz (Fig. 2e, inset).

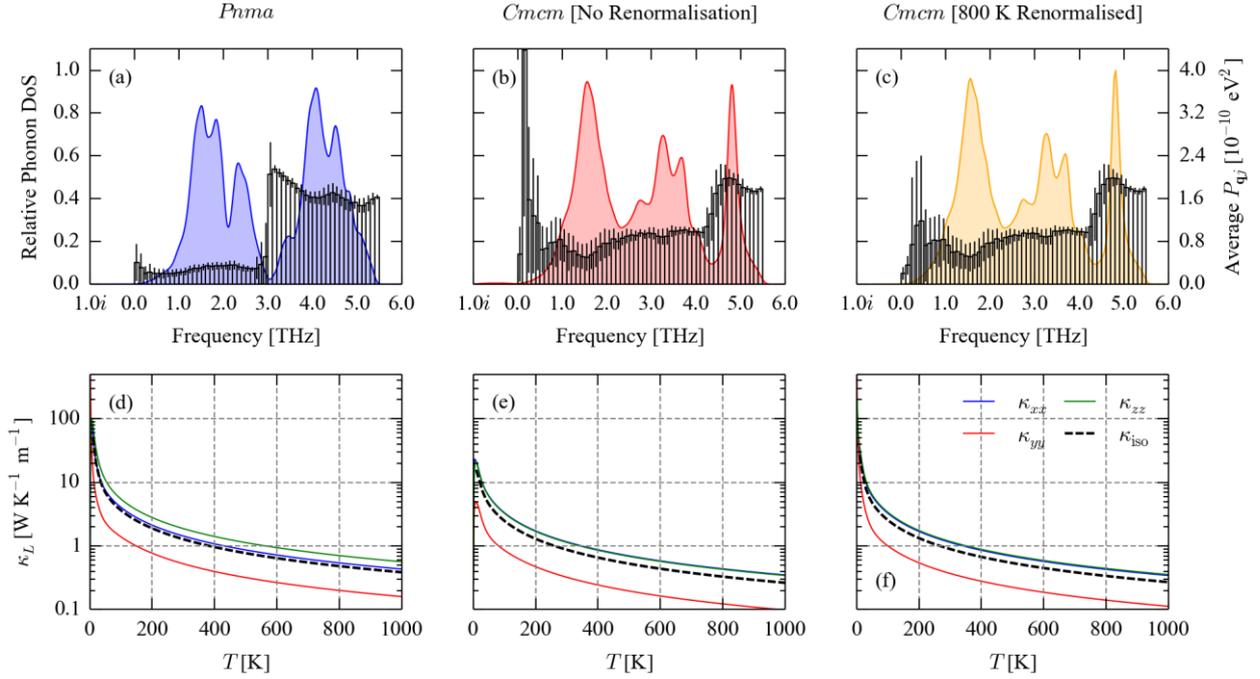

FIG. 3 (Color online) Thermal-transport properties of the *Pnma* (a, d) and *Cmcm* (b, c, e, f) phases of SnSe. Plots (a) - (c) show the averaged three-phonon interaction strengths, $P_{qj}$ (as defined in Ref. [28]), overlaid on the phonon densities of states. The error bars show the standard deviation on the average in each histogram bin. Plots (d) - (e) show the calculated lattice thermal conductivity ($\kappa_L$) along each of the three crystallographic axes, together with the isotropic average. The *Cmcm*-phase data shown in (b) and (e) was modelled without renormalization of the imaginary modes, while that shown in (c) and (f) was modelled with the second-order (harmonic) force constants corrected using the 800 K renormalized frequencies.

was also found to have a negligible impact on thermodynamic functions calculated within the harmonic approximation,[41] with a maximum difference in the high-temperature (1000 K) vibrational Helmholtz free energy of < 0.3 kJ mol$^{-1}$ per SnSe formula unit (~0.3 %; c.f. $k_B T = 8.314$ kJ mol$^{-1}$ at this temperature).

To investigate the effect of the phase transition on the thermal transport, we modelled the phonon lifetimes within the single-mode relaxation-time approximation, which were used to solve the Boltzmann transport equations. The lifetimes are calculated as the phonon self-energy within many-body perturbation theory, using the three-phonon interaction strengths $\phi_{\lambda\lambda'\lambda''}$, obtained from third-order force constants $\phi_{\alpha\beta\gamma}$ along with an expression for the conservation of energy. This method is implemented in the Phono3py code, and a detailed overview may be found in Ref. [28]. In these calculations, we did not attempt to correct the third-order force constants for the imaginary modes in the *Cmcm* phase, but instead performed post-processing for the high-temperature structure using renormalized second-order force constants.

Neglecting quasi-harmonic effects from volume expansion, $\phi_{\lambda\lambda'\lambda''}$ is temperature independent, and a histogram showing the average phonon-phonon interaction strengths, $P_{qj}$, across the phonon DoS (Fig. 3a-3c) provides a convenient means to compare the inherent "anharmonicity" of the two phases. This comparison clearly shows that the low-frequency modes in the *Cmcm* phase experience a substantially stronger interaction with other phonon modes compared to the lower-frequency branches in the *Pnma* phase. Correcting the second-order force constants reduces the interaction strength up to ~0.5 THz, but the average $P_{qj}$ up to ~3 THz remains consistently higher than in the *Pnma* phase. These calculations therefore indicate that the lattice dynamics of the high-temperature phase are inherently more anharmonic.

The calculated 300 K lattice thermal conductivity ($\kappa_L$) of the *Pnma* phase (Fig. 3d) is 1.43, 0.52 and 1.88 W m$^{-1}$ K$^{-1}$ along the $a$, $b$ and $c$ axes, respectively. These values are in reasonable agreement with measurements of ~0.7 and ~0.45 W m$^{-1}$ K$^{-1}$ along the two short and long axes,[13] and the predicted axial anisotropy is consistent with other studies.[13,16] For the *Cmcm* phase (Fig. 3e), we calculated an isotropic average $\kappa_L$ of 0.33 W m$^{-1}$ K$^{-1}$ at 800 K compared to the measured value of ~0.25 W m$^{-1}$ K$^{-1}$.[13] Given the neglect of volume expansion, higher-order anharmonicity and other potential scattering processes in the present calculations, this is again reasonable agreement. As in the *Pnma* phase, the calculations predict an axial anisotropy in $\kappa_L$, with significantly reduced transport along the long axis. With the second-order force constants corrected, we calculated an average $\kappa_L$ of 0.34 (an increase of 2.4 %), indicating that renormalization has only a small quantitative impact. Analysis of the modal contributions to the thermal transport at 300 and 800 K[41] shows that the low-frequency modes account for the bulk of the heat transport in both phases, and that the larger phonon-phonon interaction strengths in the *Cmcm* phase significantly reduce the lifetimes of these modes, thus explaining its reduced $\kappa_L$.

These findings are consistent with the mechanism proposed in Refs. [14,17]. Our calculations show that the high-temperature phase exhibits a substantial phonon softening compared to the *Pnma* phase, and that enhanced phonon-phonon interactions in the *Cmcm* phase lead to an anharmonic damping of the low-frequency modes and hence the thermal transport. Moreover, our renormalisation scheme provides a means to explore the physics of the soft modes and their effect on calculated properties, such as the thermodynamic free energy and the thermal transport, from first principles, without requiring experimental measurements of the phonon frequencies to fit to.[14]

In this regard, our scheme could easily be combined with the quasi-harmonic approximation, which would allow us to explore the effect of thermal expansion at finite temperature. This would be expected to further dampen the thermal conductivity,[24] and detailed lattice-dynamics calculations could allow the relative contributions from changes in phonon lifetimes and group velocities to be discerned.[14]

In summary, our calculations demonstrate that the low high-temperature lattice thermal conductivity of SnSe is due to anharmonic damping of the low-frequency phonon modes. We have developed a simple renormalization scheme to quantify the impact of the phonon instabilities in the high-temperature phase on properties calculated using second-order force constants, and hence shown the enhanced anharmonicity to be an inherent property of this system. This scheme may form a practical basis for studying other important classes of system with displacive instabilities, e.g. halide perovskites. From a materials-design perspective, similar anharmonic phonon dampening may occur in other systems at the boundary of a phase transition, and so this could serve as a selection criterion for identifying materials with ultra-low thermal conductivity.[14] In these materials, the poor thermal transport is a bulk property, and so the potential negative impact on electrical properties of modifications such as doping and nanostructuring may be avoided. Understanding this phenomenon may thus provide a robust design strategy for developing thermal insulators and high-performance thermoelectric materials.

We gratefully acknowledge helpful discussions with Dr J. M. Frost. JMS is supported by an EPSRC programme grant (grant no. EP/K004956/1). LAB is currently supported by the JSPS (grant no. 26.04792). JB acknowledges support from the EPSRC (grant no. EP/K016288/1). Calculations were carried out on the ARCHER supercomputer, accessed through membership of the UK HPC Materials Chemistry Consortium, which is funded by the EPSRC (grant no. EP/L000202). We also made use of the Bath University HPC cluster, which is maintained by the Bath University Computing Services, and the SiSu supercomputer at the IT Center for Science (CSC), Finland, via the Partnership for Advanced Computing in Europe (PRACE) project no. 13DECI0317/IsoSwitch.

**APPENDIX: Data-access statement**

*a.walsh@bath.ac.uk

# Anharmonicity in the High-Temperature *Cmcm* Phase of SnSe: Soft Modes and Three-Phonon Interactions

Jonathan M. Skelton, Lee A. Burton, Stephen C. Parker and Aron Walsh

*Department of Chemistry, University of Bath, Claverton Down, Bath BA2 7AY, UK*

Chang-Eun Kim and Aloysius Soon

*Department of Materials Science and Engineering, Yonsei University, Seoul 120-749, Korea*

John Buckeridge, Alexey A. Sokol and C. Richard A. Catlow

*University College London, Kathleen Lonsdale Materials Chemistry, Department of Chemistry, 20 Gordon Street, London WC1H 0AJ, United Kingdom*

Atsushi Togo and Isao Tanaka

*Elements Strategy Initiative for Structural Materials, Kyoto University, Kyoto Prefecture 606-8501, Japan*

# Electronic Supporting Information

## 1. *Ab initio* lattice-dynamics calculation protocol

First-principles calculations were carried out within the pseudopotential plane-wave density-functional theory formalism, as implemented in the Vienna *ab initio* Simulation Package (VASP) code.[1] We used projector augmented-wave (PAW) pseudopotentials,[2,3] treating the Sn 5s, 5p and 4d and the Se 4s and 4p electrons as valence states, in conjunction with the PBEsol exchange-correlation functional[4] and a plane-wave basis with a kinetic-energy cutoff of 500 eV. An 8×4×8 Monkhorst-pack **k**-point mesh[5] was used to sample the first Brillouin zone of the conventional *Pnma* and *Cmcm* cells, which was correspondingly reduced for the supercell-phonon calculations. The PAW projection was performed in reciprocal space, and non-spherical contributions to the gradient corrections inside the PAW spheres were taken into account. The electronic-structure calculations were performed within the scalar relativistic approximation, excluding spin-orbit coupling.

During geometry optimizations, a tolerance of $10^{-8}$ eV was applied during the electronic minimisation, and the ion positions and lattice parameters were optimized until the magnitude of the forces on the ions was below $10^{-2}$ eV Å$^{-1}$.

Lattice-dynamics calculations were performed with the Phonopy[6,7] and Phono3py[8] packages, which were used to obtain sets of second- and third-order force-constant matrices, respectively, *via* the supercell finite-displacement method.[9] The second-order force constants were calculated using 6×1×6 expansions of the conventional *Pnma* and *Cmcm* cells, containing 288 atoms, with a displacement step size of $10^{-2}$ Å. We found that these cells were sufficiently large to converge the shape of the phonon density of states (DoS; see Section 2, below). We also performed additional calculations on 2×6×2 and 6×6×2 supercells to obtain higher-quality dispersions along the Γ-*Y* and *S*-Γ segments of the *Cmcm* band dispersion, respectively (see Fig. 1b in the text). Due to the unfavourable scaling of the number of inequivalent two-atom displacements with supercell size, the third-order force constants were calculated from 3×1×3 supercell expansions containing 72 atoms using a step size of $3×10^{-2}$ Å. During the post processing, the phonon DoS curves were constructed by evaluating the phonon frequencies on a uniform 48×48×48 Γ-centred **q**-point grid, while the phonon lifetimes used to model the thermal transport were sampled on 16×16×16 Γ-centred grid.

The phonon DoS and thermal conductivity of the *Cmcm* phase were calculated with respect to the conventional cell, while the dispersion was calculated in the primitive basis. The transformation from the conventional to primitive cell was performed using the transformation matrix:

$$\begin{pmatrix} 1/_2 & -1/_2 & 0 \\ 1/_2 & 1/_2 & 0 \\ 0 & 0 & 1 \end{pmatrix}$$

The dispersions of the *Pnma* and *Cmcm* phases (the latter with respect to the primitive cell) were constructed by evaluating the phonon frequencies along a path connecting the high-symmetry points in the respective Brillouin zones. The reduced **q**-vectors of the symmetry points used to generate the *Pnma* and *Cmcm* dispersions are listed in Tables S1 and S2, respectively.

| Symmetry Point | Reduced **q**-Vector |
| --- | --- |
| Γ | 0.0, 0.0, 0.0 |
| X | 0.5, 0.0, 0.0 |
| S | 0.5, 0.5, 0.0 |
| Y | 0.0, 0.5, 0.0 |
| T | 0.0, 0.5, 0.5 |
| R | 0.5, 0.5, 0.5 |
| U | 0.5, 0.0, 0.5 |
| Z | 0.0, 0.0, 0.5 |

**Table S1** Reduced **q**-vectors of the high-symmetry points in the Brillouin zone of the *Pnma* phase used in the simulation of the phonon-dispersion curves.

| Symmetry Point | Reduced **q**-Vector |
| --- | --- |
| Z | 0.0, 0.0, 0.5 |
| R | 0.0, 0.5, 0.5 |
| S | 0.0, 0.5, 0.0 |
| G | 0.0, 0.0, 0.0 |
| Y | 0.5, 0.5, 0.0 |
| T | 0.5, 0.5, 0.5 |

**Table S2** Reduced **q**-vectors of the high-symmetry points in the Brillouin zone of the *Cmcm* phase used in the simulation of the phonon-dispersion curves. The coordinates are given with respect to the reciprocal lattice vectors of the primitive cell, with the transformation from the conventional to the primitive cell performed as outlined above.

## 2. Supercell-size convergence for the finite-displacement calculations

As noted in the text and in Section 1 above, we carefully converged the supercell sizes used in the finite-displacement calculations on both the *Pnma* and *Cmcm* phases. Table S3 summarises the supercell sizes considered during these convergence tests, together with the number of atoms and commensurate **q**-points in each (i.e. the number of **q**-points at which the phonon frequencies and eigenvectors can be calculated exactly for the given expansion).

In our optimised structures, the $a$ and $c$ axes are similar in length, and both are significantly shorter than the $b$ axis, which corresponds to the weakly-bonded layering direction (*Pnma*: $a = 4.367$, $b = 11.433$, $c = 4.150$ Å; *Cmcm*: $a = 4.217$, $b = 11.525$ and $c = 4.204$ Å). We therefore considered expansions of 2, 4, 6 and 8× along the $a$ and $c$ axes, together with expansions of 1 and 2× along the $b$ axis. The calculated phonon band structures and DoS curves for the *Pnma* and *Cmcm* phases with the various expansions are compared in Figs. S1 and S2, respectively. We note that an 8×2×8 expansion would require calculations on 1,024-atom supercells, and we found when attempting these that they required an impractical amount of computing resources with the tight convergence criteria needed to obtain accurate forces.

| Supercell | # Atoms | #**q**-Points |
|-----------|---------|---------------|
| 2×1×2 | 32 | 4 (8) |
| 2×2×2 | 64 | 8 (16) |
| 4×1×4 | 128 | 16 (32) |
| 4×2×4 | 256 | 32 (64) |
| 6×1×6 | 288 | 36 (72) |
| 6×2×6 | 576 | 72 (144) |
| 8×1×8 | 512 | 64 (128) |

**Table S3** List of expansions tested for supercell-size convergence, together with the number of atoms and commensurate **q**-points in each. For the *Cmcm* phase, the *C*-centring results in the conventional cell having twice the volume of the primitive cell, and a supercell expansion with a given number of commensurate points in the conventional Brillouin zone includes twice as many points in the primitive Brillouin zone; this is indicated by the numbers in brackets in the third column.

Going from a 2× to a 4× expansion along the short axes, there are noticeable changes to the shape of the DoS of both phases, while further, more subtle, changes occur when the expansions are increased to 6 and 8×. Performing a 2× expansion along the $b$ axis has comparatively little effect, particularly when larger expansions along the $a$ and $c$ axes are used, and the shape of the DoS computed with the 6×1×6 and 6×2×6 supercells is very similar, particularly for the *Cmcm* phase.

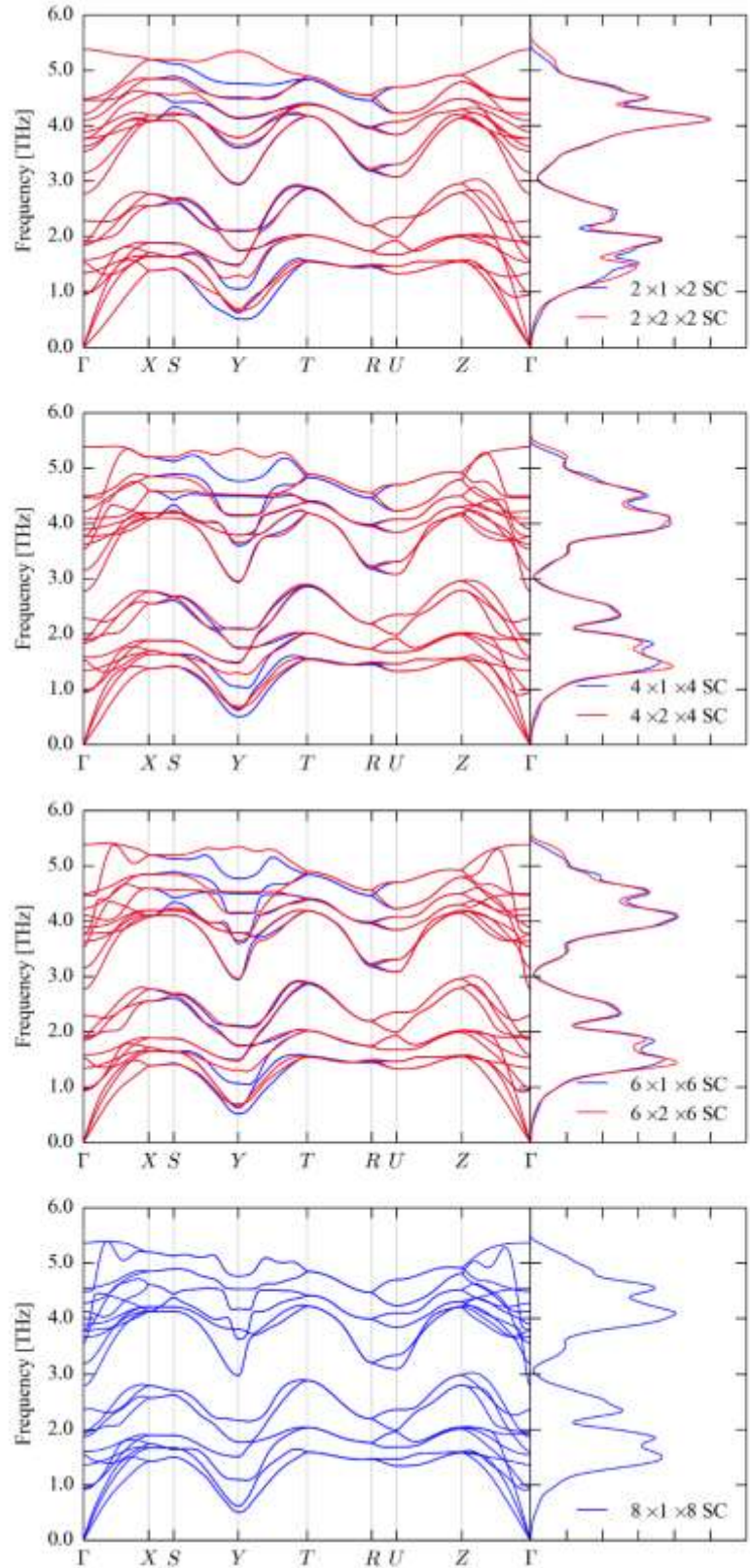

**Figure S1** Comparison of the phonon dispersion and density-of-states curves for the equilibrium structure of *Pnma* SnSe, computed using supercell expansions of 2, 4, 6 and 8× along the *a* and *c* axes, together with 1 and 2× expansions along the *b* axis, to evaluate the force-constant matrices.

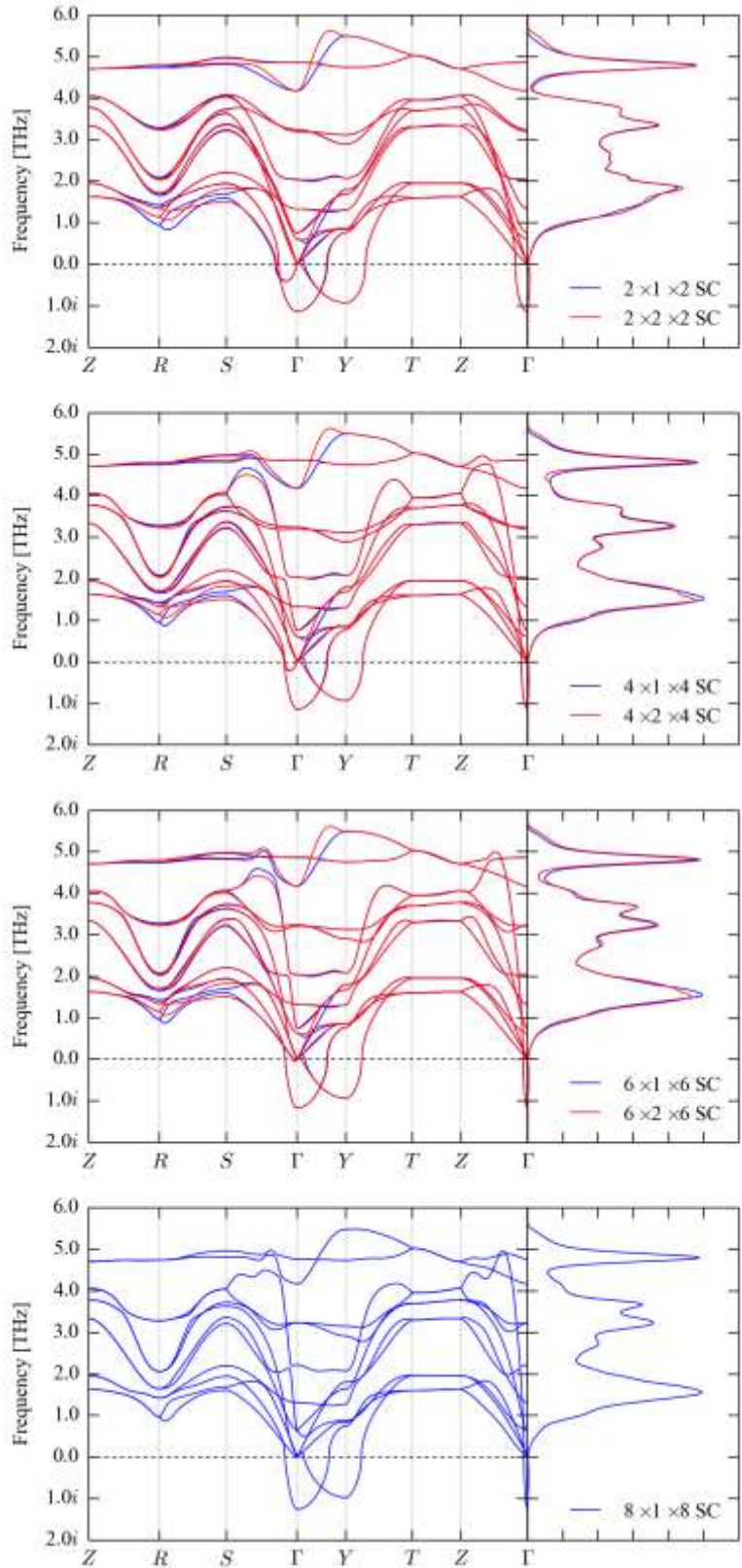

**Figure S2** Comparison of the phonon dispersion and density-of-states curves for the equilibrium structure of *Cmcm* SnSe, computed using supercell expansions of 2, 4, 6 and 8× along the *a* and *c* axes, together with 1 and 2× expansions along the *b* axis, to evaluate the force-constant matrices.

A more quantitative way to assess the supercell-size convergence is to compare the vibrational constant-volume (Helmholtz) free energies, $A_{vib}$, which are computed by summation over a grid of $\mathbf{q}$-points sampling the phonon Brillouin zone according to (eq. S1):

$$A_{vib} = -k_B T \ln \left[ \prod_\lambda \frac{\exp(-\hbar\omega_\lambda/2k_B T)}{1 - \exp(-\hbar\omega_\lambda/k_B T)} \right] \qquad (S1)$$

$k_B$ is the Boltzmann constant, $T$ is the temperature, and the product runs over the phonon modes $\lambda$ with frequencies $\omega_\lambda$.

Table S4 compares the high-temperature (1000 K) free energies of both systems computed from the calculations performed with the different supercell sizes. Taking the 2×1×2 supercell, which is a reasonable absolute minimum, as a reference, the largest deviations were found to be 0.672 and 0.345 kJ mol$^{-1}$ per SnSe formula unit for the $Pnma$ and $Cmcm$ phases, respectively, which corresponds to < 1 %. This is also an order of magnitude smaller than $k_B T$ at 1000 K (8.314 kJ mol$^{-1}$), and is at least comparable to the error one would expect from using an approximate exchange-correlation functional, suggesting that for this system the free energies are relatively insensitive to the set of force-constant matrices.

| Supercell | # $\mathbf{q}$-Points | $Pnma$ | | $Cmcm$ | |
|---|---|---|---|---|---|
| | | $A_{vib,1000K}$ [kJ mol$^{-1}$ Per F.U.] | $\Delta$ [kJ mol$^{-1}$ Per F.U.] | $A_{vib,1000K}$ [kJ mol$^{-1}$ Per F.U.] | $\Delta$ [kJ mol$^{-1}$ Per F.U.] |
| 2×1×2 | 4 | -102.070 | 0.000 | -104.800 | 0.000 |
| 2×2×2 | 8 | -101.721 | 0.349 | -104.614 | 0.186 |
| 4×1×4 | 16 | -101.863 | 0.208 | -105.072 | -0.272 |
| 4×2×4 | 32 | -101.666 | 0.404 | -104.924 | -0.124 |
| 6×1×6 | 36 | -101.796 | 0.274 | -105.073 | -0.273 |
| 6×2×6 | 72 | -101.620 | 0.450 | -104.930 | -0.130 |
| 8×1×8 | 64 | -101.398 | 0.672 | -105.145 | -0.345 |

**Table S4** High-temperature (1000 K) constant-volume (Helmholtz) vibrational free energies, $A_{vib,1000K}$, of $Pnma$ and $Cmcm$ SnSe, computed from harmonic phonon calculations performed on the equilibrium structures with the different supercell sizes listed in Table S3.

One potential issue with $N$×1×$N$ expansions is that while the $\Gamma$ and $Y$ points in the $Cmcm$ primitive cell, where the principal soft modes occur, are both commensurate, these supercells contain no commensurate points along the line between them. To check the convergence of the $\Gamma$-$Y$ segment of the dispersion, we therefore performed a systematic set of calculations on 2×$N$×2 expansions, with $N$ = 1, 2, 4, 6 and 8 (Fig. S3).

As noted in the text, these tests confirmed that the soft modes lie along the line between $\Gamma$ and $Y$, rather than being localised to the symmetry points. However, we found that the smaller expansions, including $N$ = 1, reproduced the shape of the dispersion of the soft modes along this segment obtained from the larger calculations. The main effect of larger values of $N$ is seen in the dispersion of the highest-energy optic branch, introducing oscillations which may be due to our not including non-analytical corrections for LO/TO splitting in these calculations. However, the effect on the phonon DoS is subtle, confirming that including longer-range interactions along the $a$ and $b$ directions are more important for reproducing the distribution of phonon frequencies.

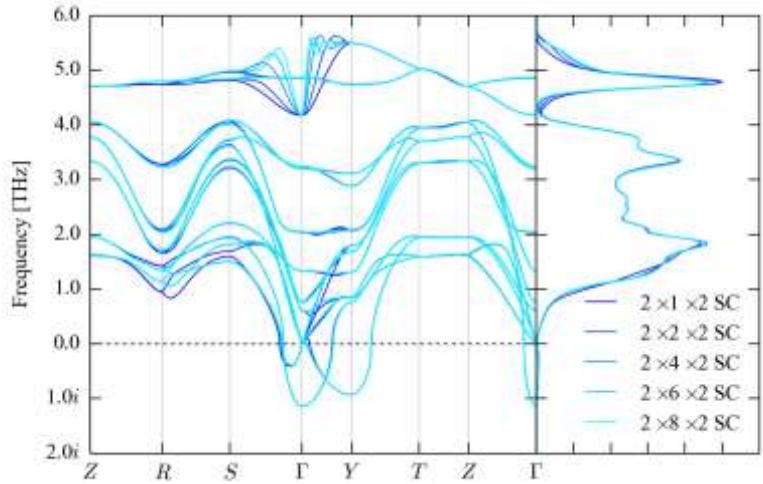

**Figure S3** Comparison of the phonon dispersion and density of states curves for *Cmcm* SnSe obtained from 2×*N*×2 expansions of the conventional cell with *N* = 1, 2, 4, 6 and 8. Note that the additional soft mode along the Γ-*S* segment is an interpolation artefact from the small expansion in the *a* and *c* directions, and is not seen in larger supercells expanded along these directions (c.f. Fig. S2).

Based on these considerations, we selected 6×1×6 expansions for our "production" calculations, as a balance between computational cost and accuracy.

**3. Numerical integral of the phonon densities of states**

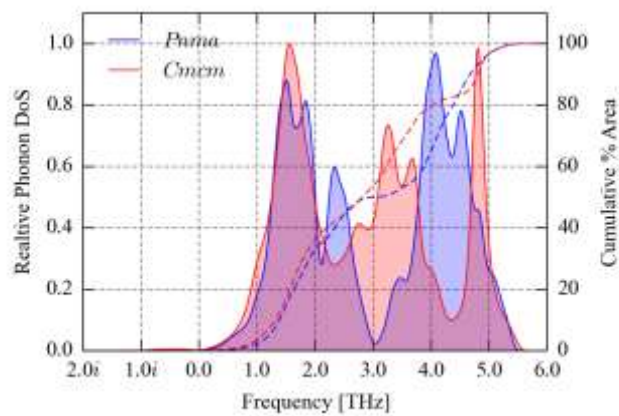

**Figure S4** Comparison of the phonon density-of-states (DoS) curves for the equilibrium *Pnma* (blue) and *Cmcm* (red) SnSe structures, showing a red shift of the higher-frequency modes (with frequencies > 3 THz) in the latter compared to the former. The two dashed lines show the cumulative area (number of states) under each curve as a function of frequency, allowing the shift to be quantified. As stated in the text, whereas there are a similar number of states below 3 THz in both phases, between 3 - 4 THz there are ~15 % more states in the *Cmcm* phase DoS than in the corresponding *Pnma* DoS.

## 4. Eigenvectors of the imaginary modes in the equilibrium *Cmcm* structure

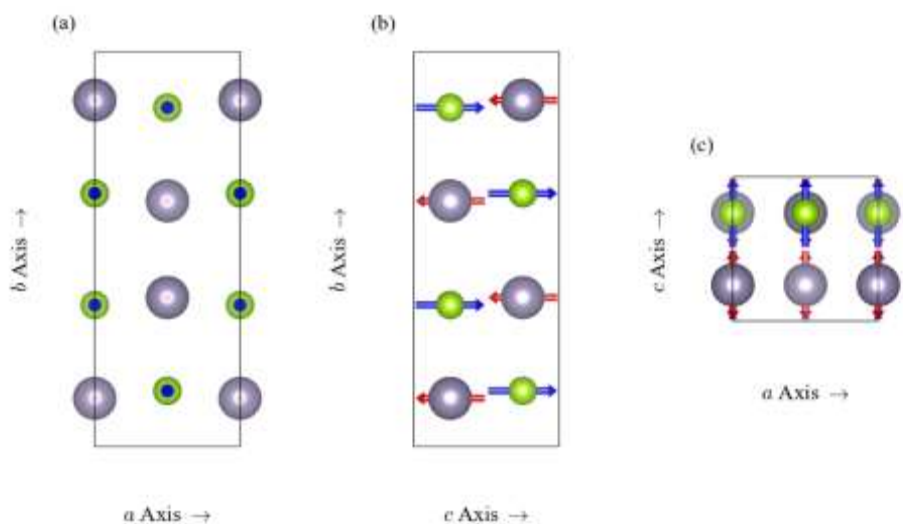

**Figure S5** Eigenvectors of the imaginary mode with $v = 1.137i$ THz in the conventional cell of the equilibrium *Cmcm* SnSe structure. This corresponds to the imaginary mode at $\Gamma$ in the primitive cell (labelled $\lambda_1$ in Fig. 1b in the text). The eigenvectors are shown in the $ab$ (a), $bc$ (b) and $ac$ (c) planes. The arrows indicate the direction, but not the magnitude, of the atomic displacements. The images were prepared with the VESTA software.[10]

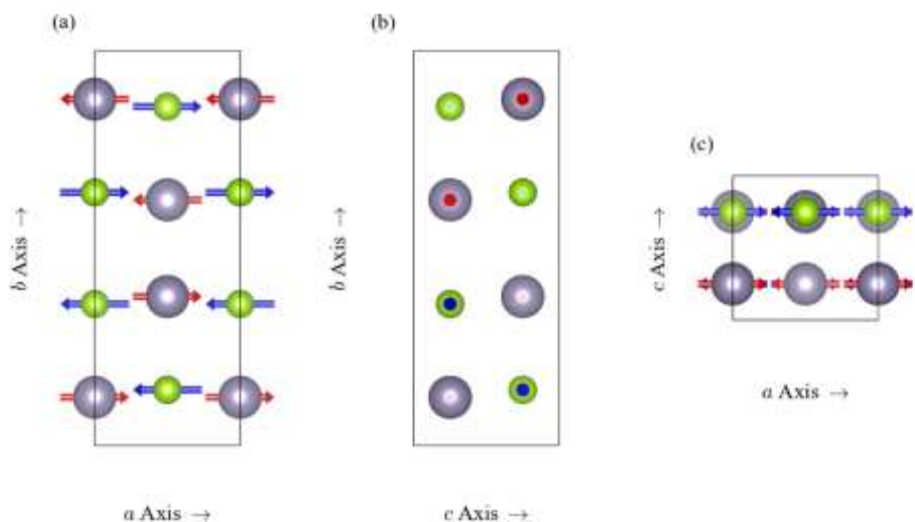

**Figure S6** Eigenvectors of the imaginary mode with $v = 0.928i$ THz in the conventional cell of the equilibrium *Cmcm* SnSe structure. This corresponds to the imaginary mode at $Y$ in the primitive cell (labelled $\lambda_2$ in Fig. 1b in the text). The eigenvectors are shown in the $ab$ (a), $bc$ (b) and $ac$ (c) planes. The arrows indicate the direction, but not the magnitude, of the atomic displacements. The images were prepared with the VESTA software.[10]

## 5. Volume dependence of the *Cmcm* phonon dispersion and density of states

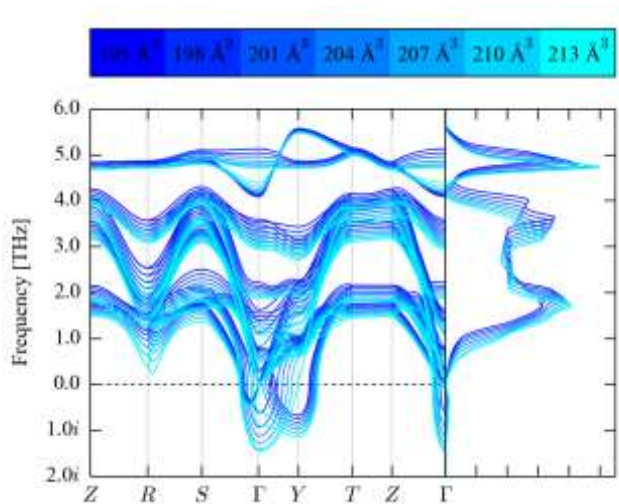

**Figure S7** Volume dependence of the phonon dispersion and density of states of *Cmcm* SnSe, computed using 2×1×2 supercells to evaluate the force-constant matrices. The lines are colour coded from blue to cyan, corresponding to compressions and expansions, respectively, about the equilibrium volume. The smallest volume corresponds to a hydrostatic pressure of 2.17 GPa. As noted in the text, the imaginary modes at the Γ and *Y* symmetry points soften under expansion and harden under compression, but persist over the range of volumes examined. Note that, as in Fig. S3, the additional soft mode along the Γ-*S* segment is an interpolation artefact from the small expansion along the *a* and *c* directions, and is not seen in supercells with larger expansions along these directions.

## 6. Imaginary-mode mapping and renormalization scheme

The potential-energy surfaces along the two imaginary modes in the equilibrium *Cmcm* structure (Figs. 1e and 2a/c in the text) were mapped by performing single-point total-energy calculations on a series of structures of the conventional cell with atomic displacements along the mode eigenvectors over a range of amplitudes "frozen in". Given the mode eigenvectors $\boldsymbol{W}_\lambda$ of a set of phonon modes $\lambda$, together with the corresponding normal-mode coordinates (amplitudes) $Q_\lambda$, the displacement $\boldsymbol{u}_{j,l}$ of the $j$th atom in the $l$th unit cell can be calculated according to:

$$\boldsymbol{u}_{j,l} = \frac{1}{\sqrt{n_a m_j}} \text{Re} \left[ \sum_\lambda Q_\lambda \boldsymbol{W}_{\lambda,j} \exp(-i\mathbf{q}.\boldsymbol{r}_{jl}) \right] \tag{S2}$$

where $\boldsymbol{W}_{\lambda,j}$ is the component of $\boldsymbol{W}_\lambda$ on atom $j$, $m_j$ are the atomic masses, $n_a$ is the number of atoms in the supercell used to model the displacement, $\mathbf{q}$ is the phonon wavevector and $\boldsymbol{r}_{jl}$ are the positions of the atoms. We note that $Q_\lambda$ absorbs the time dependence of the position.

As noted in the text, the mode at $Y$ in the Brillouin zone of the *Cmcm* primitive cell folds to $\Gamma$ in the reciprocal space of the conventional cell. In the special case of mapping $\Gamma$-point modes, Eq. S2 can be simplified to:

$$\boldsymbol{u}_{j,l} = \frac{1}{\sqrt{n_a m_j}} \sum_\lambda Q_\lambda \boldsymbol{W}_{\lambda,j} \tag{S3}$$

In the present study, the potential-energy surfaces were mapped for values of $Q_1 = \pm 30$ and $Q_2 = \pm 45$ amu$^{\frac{1}{2}}$ Å in steps of 0.5 amu$^{\frac{1}{2}}$ Å. The resulting $U(Q_i)$ curves were fitted to 20-power polynomial functions, which were used as input to a program written to solve a 1D Schrödinger equation (1D SE) for the potential to obtain the energy levels (eigenvalues) within the double wells (Figs. 2a/2c in the text).

Each polynomial was evaluated on a 1D grid of $Q$ points, and the resulting potential substituted into the 1D SE and solved for the eigenvalues and eigenvectors by means of a Fourier transform followed by matrix diagonalization using the EISPACK routines.[11] This procedure and its advantages compared to solutions in real space are explained in detail in Ref. [12], and we have verified its efficiency against more conventional "shooting method" approaches.

The eigenvalues, $E_i$, are used to determine the partition function, $Z$, via:

$$Z(T) = \sum_i e^{-E_i/k_B T} \tag{S4}$$

where $k_B$ is the Boltzmann constant and $T$ is the temperature. The number of terms included is chosen so that the addition of the final term changes $Z$ by less than $10^{-6}$. By setting this expression for $Z$ equal to the harmonic partition function:

$$Z_{harm}(T) = \sum_n e^{-\left(n+\frac{1}{2}\right)\hbar\tilde{\omega}(T)/k_B T} \tag{S5}$$

where $\tilde{\omega}$ is the effective harmonic frequency, we can derive that:

$$\tilde{\omega}(T) = \frac{2k_B T}{\hbar} \sinh^{-1}\left(\frac{1}{2Z(T)}\right) \tag{S6}$$

which provides an effective renormalized harmonic frequency for the mode at a target temperature $T$ that reproduces its contribution to the thermodynamic partition function (Figs. 2b/2d in the text).

The extent of the potential in $Q$ and the grid density were both carefully converged, with a basis of 512 grid points and $Q_1 = \pm 25$ and $Q_2 = \pm 35$ amu$^{\frac{1}{2}}$ Å, in both cases extending to ~1 eV in energy above the average structure, found to be sufficient to obtain renormalized frequencies converged to $< 10^{-2}$ THz.

These effective frequencies were used to adjust the force constants using the Python API exposed by the Phonopy code.[6,7] The original force/displacement sets and corresponding force constants were used to calculate the phonon frequencies and eigenvectors in the conventional cell at the **q**-points commensurate with our chosen supercell expansions. The frequencies of the imaginary modes at the $\Gamma$ point were then set to the calculated effective (real) ones, and the corrected dynamical matrices back-transformed to a new set of force constants to be used in subsequent post-processing steps.

A set of scripts for setting up and post-processing the displacement-mapping calculations, source code and example input files for the 1D SE solver, and a script for patching the force constants using the Phonopy API, are available as additional data (see the Appendix in the main text).

## 7. Effect of renormalisation on thermodynamic functions

To quantify the effect of renormalizing the imaginary modes in the equilibrium *Cmcm* structure on the thermodynamic functions derived within the harmonic approximation, we compared the temperature dependence of four quantities, *viz.* the constant-volume vibrational (Helmholtz) free energy, $A_{vib}$, the constant-volume heat capacity, $C_V$, the vibrational internal energy, $U_{vib}$, and the vibrational entropy, $S_{vib}$, computed with and without renormalizing the imaginary modes. We compared the original curves, i.e. computed without renormalization, to those computed with the imaginary modes renormalized to three constant values, *viz.* the effective 0, 300 and 1000 K frequencies calculated within our scheme, and to the renormalized frequency at each temperature-sampling point (see Figs. 2b/2d in the text). The results are shown in Fig. S8, and the calculated vibrational zero-point energies ($U_{ZP}$) and 1000 K free energies ($A_{vib,1000K}$) are collected in Table S5.

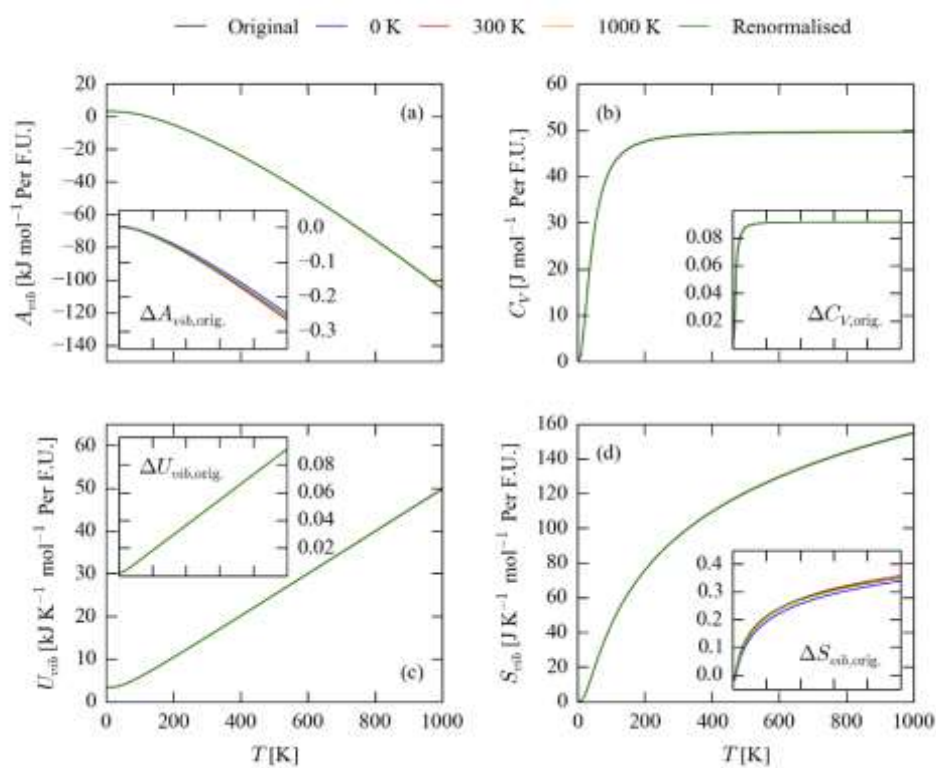

**Figure S8** Effect of imaginary-mode renormalization on the temperature-dependent constant-volume vibrational (Helmholtz) free energy ($A_{vib}$; a), constant-volume heat capacity ($C_V$; b), vibrational internal energy ($U_{vib}$; c) and vibrational entropy ($S_{vib}$; d). Each plot compares the thermodynamic functions computed using the original phonon frequencies (i.e. without renormalisation; black) and with the imaginary modes renormalized to the (constant) 0 (blue), 300 (red) and 1000 K (orange) frequencies. The green curves were computed by calculating the thermodynamic functions at each temperature-sampling point, i.e. with the imaginary modes renormalized to the corresponding frequencies in Figs. 2b/2d in the text. The inset plots show the differences between the four sets of functions computed with the various renormalization methods and the functions calculated without renormalization.

| | [THz] | | [kJ mol⁻¹ Per SnSe F.U.] | |
|---|---|---|---|---|
| Renormalization | $\tilde{v}_1$ | $\tilde{v}_2$ | $U_{ZP}$ | $A_{vib,1000K}$ |
| None | $1.137i$ | $0.928i$ | 3.388 | -105.073 |
| 0 K | 0.537 | 0.445 | 3.391 | -105.321 |
| 300 K | 0.346 | 0.233 | 3.390 | -105.340 |
| 1000 K | 0.482 | 0.316 | 3.391 | -105.332 |
| *T* dep. | - | - | 3.391 | -105.332 |

**Table S5** Calculated vibrational zero-point energies ($U_{ZP}$) and 1000 K constant-volume (Helmholtz) free energies ($A_{vib,1000K}$) without renormalization of the imaginary modes, and with the imaginary modes renormalized to three constant values, *viz.* the calculated 0, 300 and 1000 K frequencies, and to the calculated frequency at each temperature-sampling point (see Figs. 2b/2d in the text). For the first four rows, the original/renormalized frequencies of the two imaginary modes marked in Fig. 1b in the text ($\tilde{v}_{\lambda=1,2}$) that were used are listed. The $U_{ZP}$ and $A_{vib,1000K}$ values for the temperature-dependent renormalization are the same as those for the 0 and 1000 K constant renormalized frequencies, respectively.

The data in Table S5 suggests that the renormalization has a minimal impact on the free energy, leading to differences in the zero-point energy on the order of $10^{-3}$ kJ mol⁻¹ per SnSe formula unit, and to differences in the high-temperature free energy of < 0.3 kJ mol⁻¹ (~0.3 %). This is an order of magnitude smaller than $k_BT$ at 1000 K (8.314 kJ mol⁻¹). From the comparison in Fig. S8, this difference occurs predominantly as a result of an increase in $S_{vib}$, although there is also a small increase in $U_{vib}$, which is roughly the same for all four renormalized curves.

## 8. Detailed analysis of the thermal transport in the *Pnma* and *Cmcm* phases

To investigate the origin of the reduction in lattice thermal conductivity ($\kappa_L$) between the low-temperature *Pnma* and high-temperature *Cmcm* phases, we performed a detailed analysis of the modal contributions to the room-temperature (300 K) and 800 K lattice thermal conductivities of the two phases.

A comparison of the cumulative lattice thermal conductivity as a function of phonon frequency for the *Pnma* phase at 300 K against the phonon DoS (Fig. S9a) indicates that ~70 % of the heat transport is due to the lower-frequency modes (with frequencies < ~3 THz). This rises to around 80 % in the *Cmcm* phase (Fig. S9b), which can be partly explained by the red shift of some of the high-frequency modes visible in Fig. S4 (see Section 3, above). Renormalizing the soft modes has little effect on the shape of the DoS nor the cumulative thermal conductivity distribution (Fig. S9c). A similar comparison at 800 K (Fig. S13) yields a practically identical analysis.

Within the single-mode relaxation-time approximation, the macroscopic thermal-conductivity tensor is obtained as a summation of modal contributions according to:[8]

$$\boldsymbol{\kappa}_L = \frac{1}{NV_0} \sum_\lambda C_\lambda \boldsymbol{v}_{g,\lambda} \otimes \boldsymbol{v}_{g,\lambda} \tau_\lambda \tag{S7}$$

where $N$ is the number of unit cells in the crystal (equivalently, the number of reciprocal-space grid points used to sample the lifetimes), $V_0$ is the unit-cell volume, $C_\lambda$ are the modal heat capacities, $\boldsymbol{v}_{g,\lambda}$ are the mode group velocities, and $\tau_\lambda$ are the mode lifetimes. The group velocities and modal heat capacities are calculable within the harmonic approximation according to:

$$C_\lambda = \sum_\lambda k_B \left(\frac{\hbar\omega_\lambda}{k_B T}\right)^2 \frac{\exp(\hbar\omega_\lambda/k_B T)}{[\exp(\hbar\omega_\lambda/k_B T - 1)]^2} \tag{S8}$$

$$\boldsymbol{v}_{g,\lambda} = \frac{\partial \omega_\lambda}{\partial \mathbf{q}} \tag{S9}$$

The procedure for calculating lifetimes followed in this work is explained in detail in Ref. [8]. The lifetimes are derived from three-phonon interaction strengths, $\phi_{\lambda\lambda'\lambda''}$, together with and an expression for the conservation of energy. $\phi_{\lambda\lambda'\lambda''}$ are calculated according to:

$$\phi_{\lambda\lambda'\lambda''} = \frac{1}{\sqrt{N}} \frac{1}{3!} \sum_{jj'j''} \sum_{\alpha\beta\gamma} W_\alpha(\lambda,j) W_\beta(\lambda',j') W_\gamma(\lambda'',j'') \sqrt{\frac{\hbar}{2m_j\omega_\lambda}} \sqrt{\frac{\hbar}{2m_{j'}\omega_{\lambda'}}} \sqrt{\frac{\hbar}{2m_{j''}\omega_{\lambda''}}}$$
$$\times \sum_{l'l''} \phi_{\alpha\beta\gamma}(j0,j'l',j''l'') e^{i\mathbf{q}'.[r(j'l')-r(j0)]} e^{i\mathbf{q}''.[r(j'l'')-r(j0)]} e^{i(\mathbf{q}+\mathbf{q}'+\mathbf{q}'').r(j0)} \Delta(\mathbf{q}+\mathbf{q}'+\mathbf{q}'') \tag{S10}$$

As in Eqs. S2/S3, the indices $j$ and $l$ label atoms and unit cells, respectively, and $\alpha$, $\beta$, $\gamma$ are the Cartesian directions. $\phi_{\alpha\beta\gamma}$ are the third-order force-constant matrices, $\omega_\lambda$ and $\mathbf{q}$ are the phonon frequencies and wavevectors, respectively, $W_\alpha$ are the components of the phonon eigenvectors, and $r(jl)$ are the atom positions. The delta function $\Delta(\mathbf{q}+\mathbf{q}'+\mathbf{q}'')$ is unity when the sum of the wavevectors is a reciprocal lattice vector, and zero otherwise, which enforces conservation of momentum. $\phi_{\lambda\lambda'\lambda''}$ are used to calculate the imaginary part of the phonon self-energy, $\Gamma_\lambda(\omega)$, using the many-body perturbation theory result (overleaf):

$$\Gamma_\lambda(\omega) = \frac{18\pi}{\hbar^2} \sum_{\lambda'\lambda''} |\phi_{-\lambda\lambda'\lambda''}|^2 \{(n_{\lambda'} + n_{\lambda''} + 1)\delta(\omega - \omega_{\lambda'} - \omega_{\lambda''}) $$
$$ + (n_{\lambda'} - n_{\lambda''})[\delta(\omega + \omega_{\lambda'} - \omega_{\lambda''}) - \delta(\omega - \omega_{\lambda'} + \omega_{\lambda''})]\} \quad \text{(S11)}$$

where $n_\lambda$ are the phonon occupation numbers, and the delta functions enforce conservation of energy. Assuming three-phonon processes to be the dominant scattering effect, $\Gamma_\lambda(\omega)$ are related to the phonon lifetimes according to:

$$\tau_\lambda = \frac{1}{2\Gamma_\lambda(\omega_\lambda)} \quad \text{(S12)}$$

where $2\Gamma_\lambda(\omega_\lambda)$ are the full-width at half-maxima of the phonon lines.

To analyse the thermal transport in more detail, we compared the spread of the (isotropically-averaged) modal contributions to the thermal conductivity, $\kappa_\lambda$, as a function of frequency against the spreads of the group velocity norms, $|v_{g,\lambda}|$, lifetimes, $\tau_\lambda$, and average three-phonon interaction strengths, $P_{\mathbf{q}j,\lambda}$. Following Ref. [8], $P_{\mathbf{q}j,\lambda}$ are defined here as:

$$P_{\mathbf{q}j,\lambda} = \frac{1}{(3n_a)^2} \sum_{\lambda'\lambda''} |\phi_{\lambda\lambda'\lambda''}|^2 \quad \text{(S13)}$$

where $n_a$ is the number of atoms in the unit cell, and $3n_a$ is thus the number of phonon bands at each $\mathbf{q}$-point.

Figs. S10 and S11 give a breakdown of the modal contributions to the room-temperature thermal conductivity of the *Pnma* and *Cmcm* phases. In the *Pnma* phase, the dominant contribution is clearly from long-lived low-frequency modes, while higher-frequency phonons with larger group velocities account for a secondary contribution. The lifetimes of the low-frequency modes are significantly reduced in the high-temperature phase, and mid-frequency phonons with large group velocities make a proportionately higher contribution to the bulk thermal transport. A corresponding analysis of the 800 K thermal conductivity (Figs. S14, S15) shows a very similar pattern, but with a marked reduction in the phonon lifetimes of both phases, particularly of the low-frequency modes.

Finally, we also analysed the contributions of modes with different mean-free paths (MFPs) to the room-temperature thermal conductivity (Fig. S12; the MFP is calculated as $\text{MFP}_\lambda = |v_{g,\lambda}|\tau_\lambda$). We found that phonons with MFPs below ~1 nm make a very small contribution to the overall heat transport, while the longest phonon MFP observed in either phase was ~1 μm. The distribution of cumulative lattice thermal conductivity as a function of increasing phonon MFP is skewed more towards longer paths in the low-temperature *Pnma* phase than in the high-temperature *Cmcm* system, although in both around 50 % of the thermal conductivity is through phonons with MFPs < 10-20 nm. Again, renormalization of the soft modes in the high-temperature phase makes little difference to the spread of the $\kappa_\lambda$/MFPs, and has a negligible effect on the shape of the thermal-conductivity distribution. At 800 K (Fig. S16), the distributions maintain a similar overall shape, but are skewed towards phonons with shorter MFPs, as would be expected given the reduction in lifetimes evident from comparing Figs. S10/11 and S14/15.

As discussed in the text, these analyses indicate that the comparative reduction in the lifetimes of the low-frequency phonon modes in the *Cmcm* phase is primarily responsible for its lower lattice thermal conductivity. This can be ascribed to the higher average three-phonon interaction strength experienced by modes in the lower-frequency part of the DoS in the high-temperature phase (Fig. 3 in the text), which belies an inherently higher anharmonicity.

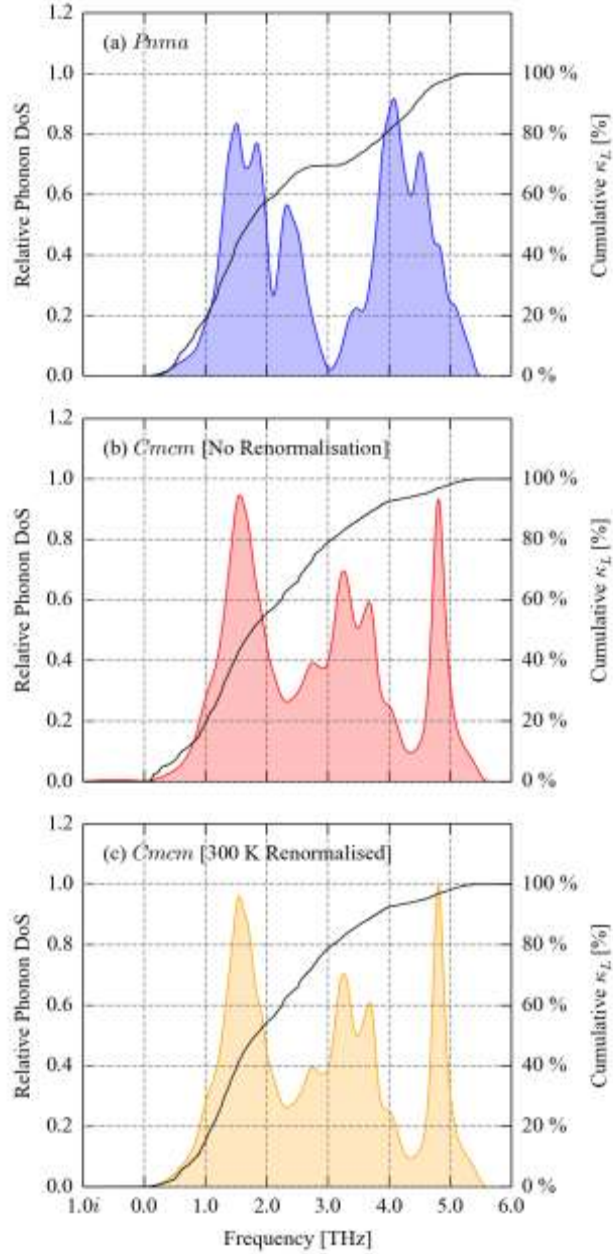

**Figure S9** Cumulative isotropically-averaged lattice thermal conductivity, $\kappa_L$, at 300 K, as a function of phonon frequency, of the *Pnma* (a) and *Cmcm* (b, c) phases of SnSe, overlaid on the corresponding phonon density of states curves. The data for the *Cmcm* phase in plot (b) is from simulations performed without renormalization of the two soft modes, while the results in plot (c) are obtained with the modes renormalized to the calculated 300 K frequencies ($\tilde{v}_1 = 0.346$ and $\tilde{v}_2 = 0.233$ THz).

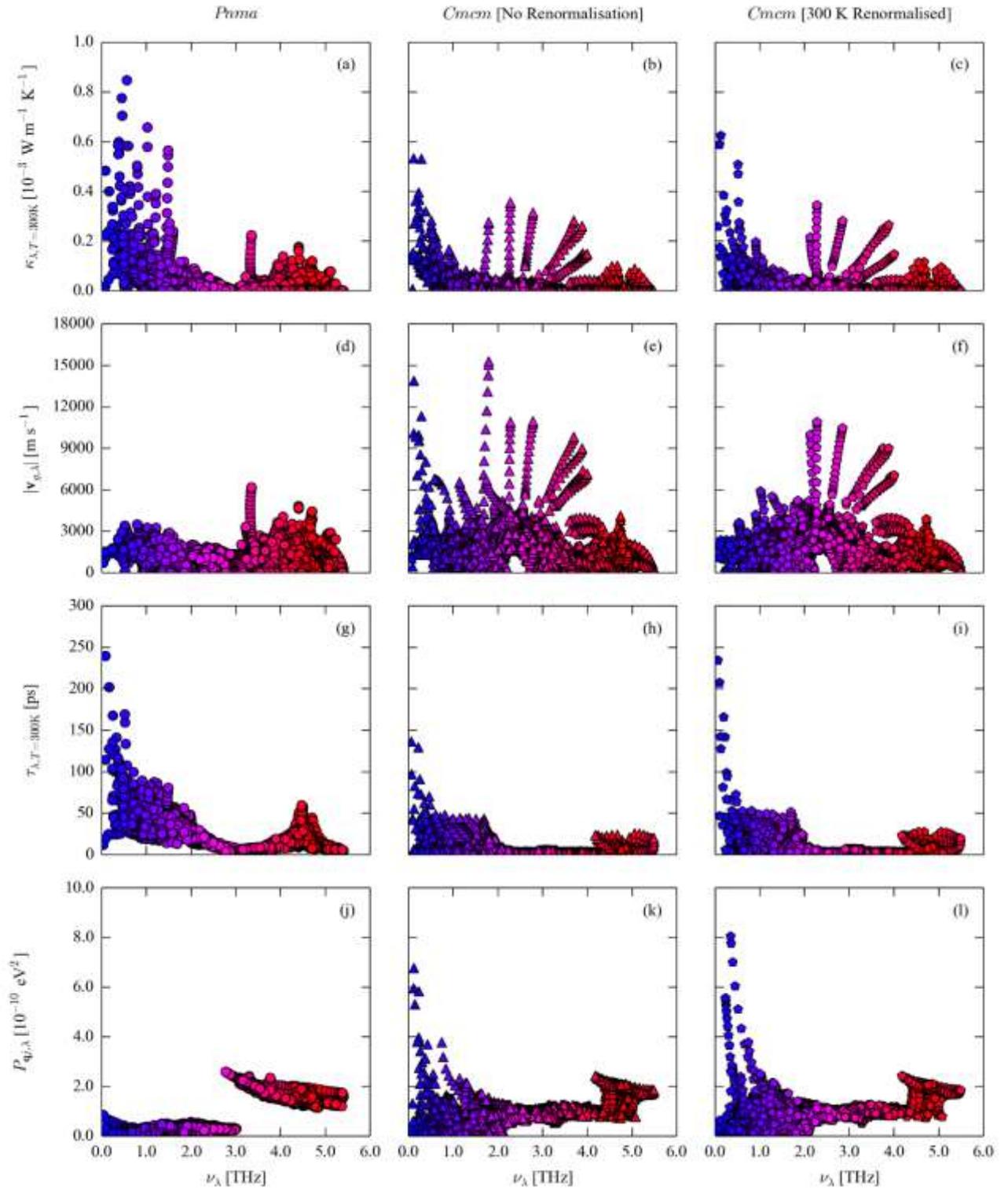

**Figure S10** Comparison of the isotropically-averaged modal contributions to the lattice thermal conductivity, $\kappa_\lambda$, of SnSe at 300 K (a-c) to the mode group velocities, $|\boldsymbol{v}_{g,\lambda}|$ (d-f), lifetimes, $\tau_\lambda$ (g-i), and averaged three-phonon interaction strengths, $P_{\mathbf{q}j,\lambda}$ (j-l; see Eq. S13). The left column shows data for the *Pnma* phase (a, d, g, j), while the other two show data for the *Cmcm* phase without renormalization of the two imaginary modes (centre; b, e, h, k) and with the modes renormalized to the calculated 300 K frequencies (right; c, f, i, l; $\tilde{v}_1 = 0.346$ and $\tilde{v}_2 = 0.233$ THz).

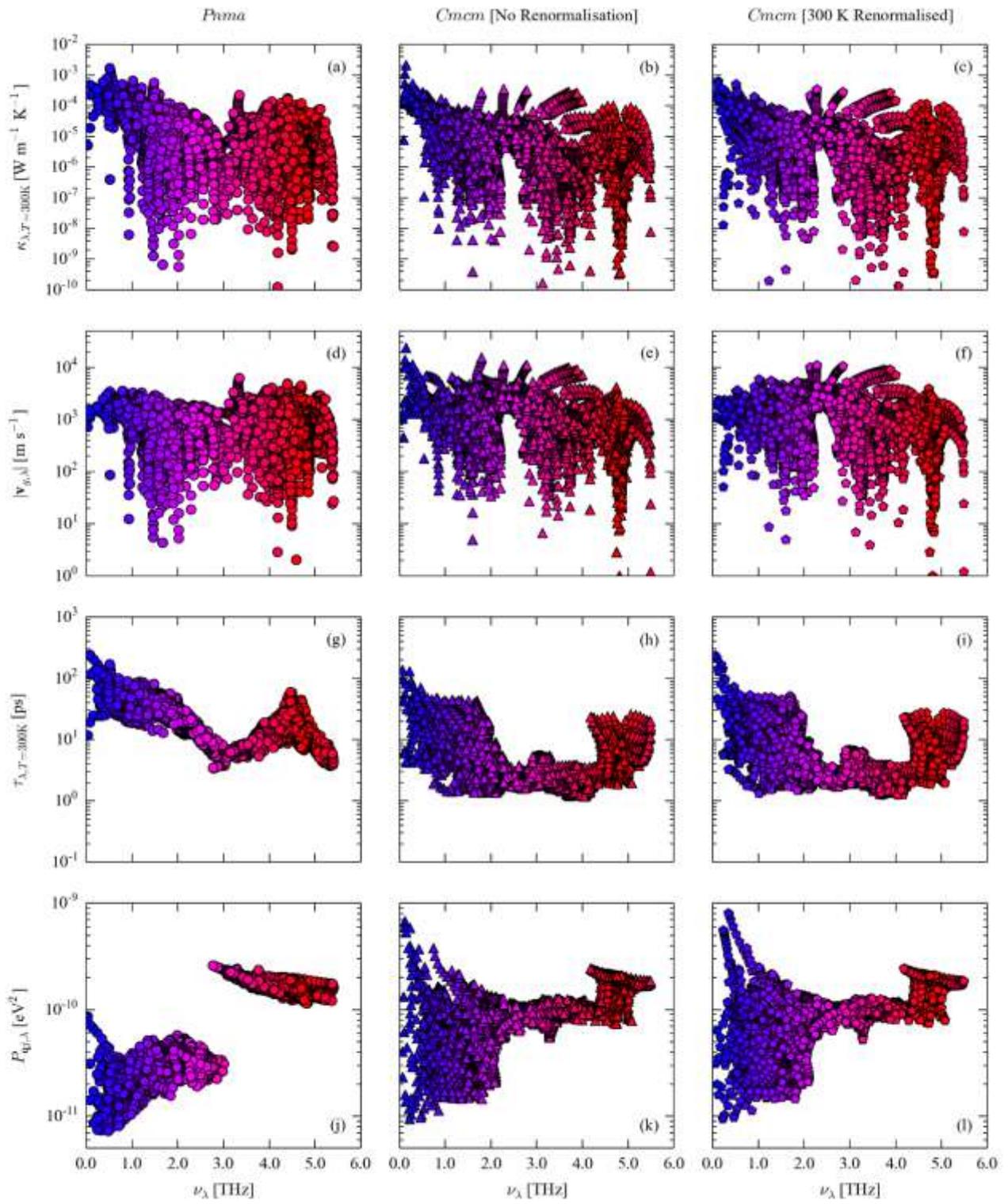

**Figure S11** Same data as in Fig. S10, plotted on a logarithmic scale.

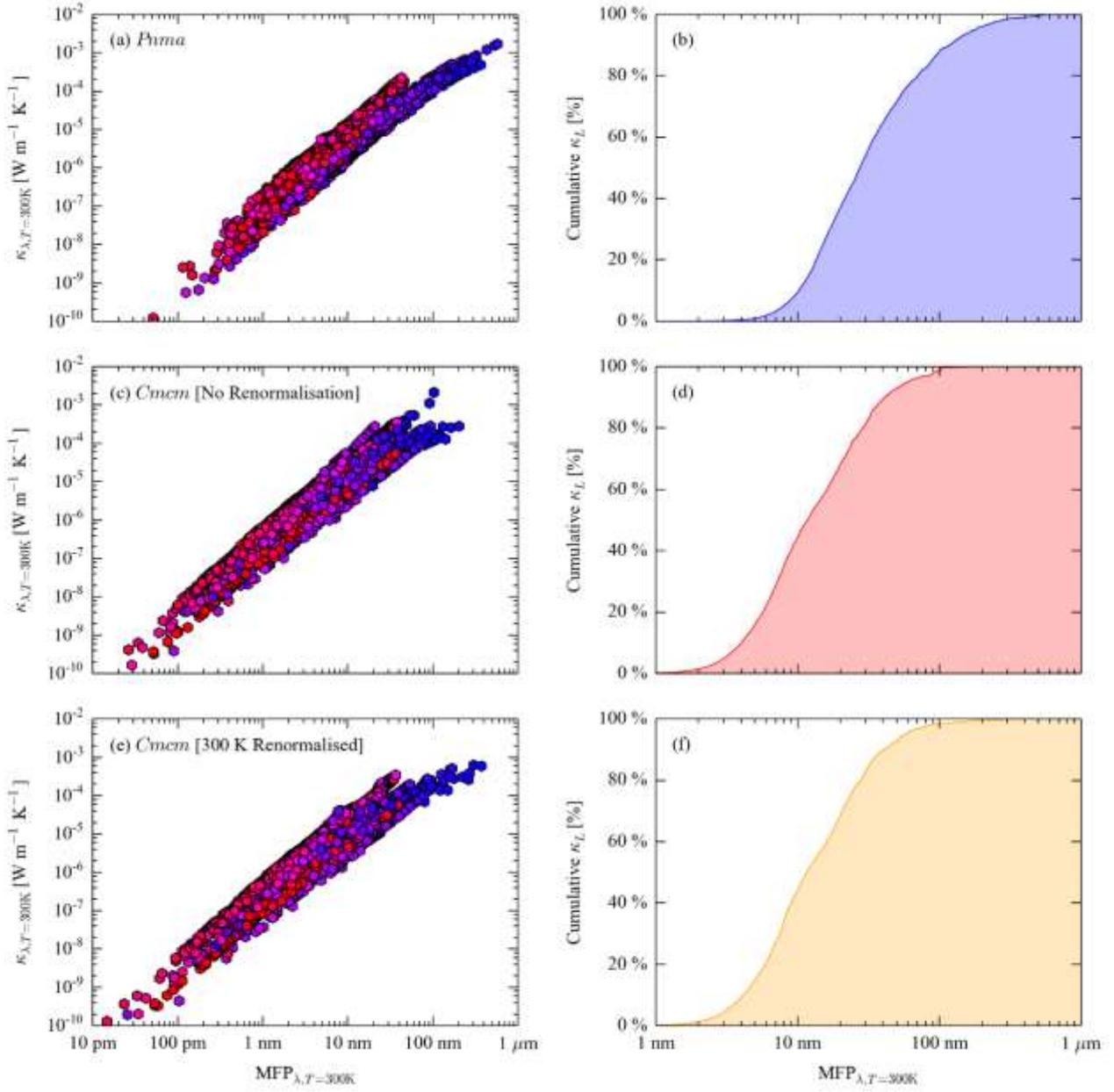

**Figure S12** Contributions of modes with different mean-free paths (MFPs) to the isotropically-averaged 300 K lattice thermal conductivity (a, c, e), together with the cumulative thermal conductivity over MFPs between 1 nm and 1 μm (b, d, f), for *Pnma* (a, b) and *Cmcm* (c-f) SnSe. For the latter high-temperature phase, two sets of data are shown: one calculated without renormalization of the two soft modes (c, d), and one with the modes renormalized to the calculated effective 300 K frequencies (e, f; $\tilde{\nu}_1 = 0.346$ and $\tilde{\nu}_2 = 0.233$ THz).

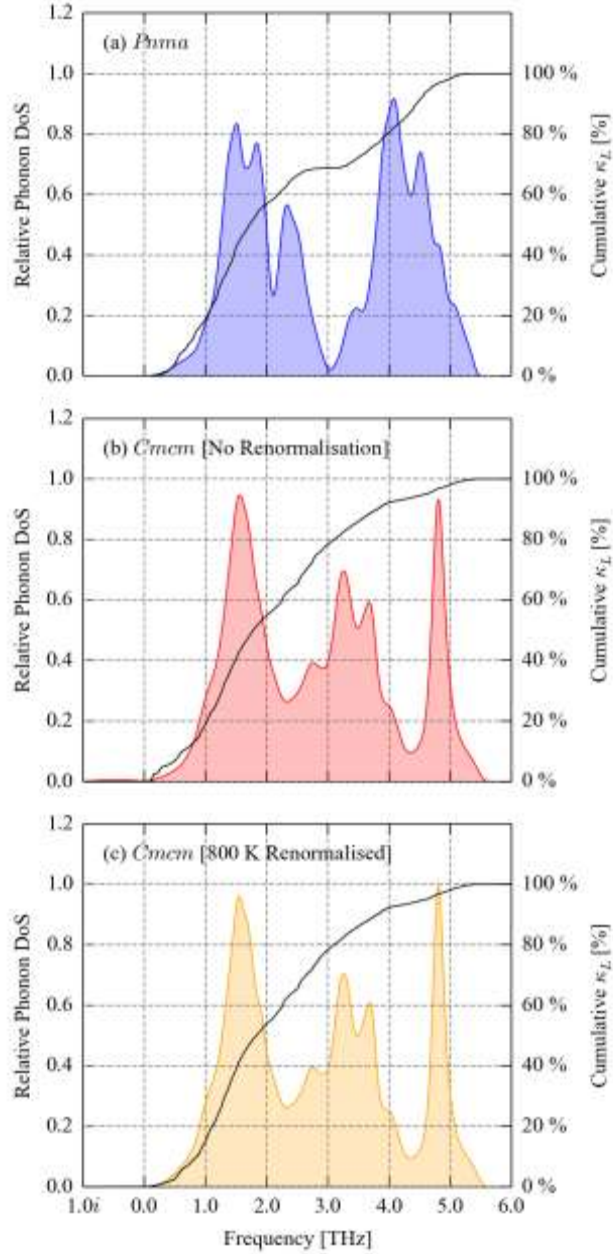

**Figure S13** Cumulative isotropically-averaged lattice thermal conductivity, $\kappa_L$, at 800 K, as a function of phonon frequency, of the *Pnma* (a) and *Cmcm* (b, c) phases of SnSe, overlaid on the corresponding phonon density of states curves. The data for the *Cmcm* phase in plot (b) is from simulations performed without renormalization of the two soft modes, while the results in plot (c) are obtained with the modes renormalized to the calculated 800 K frequencies ($\tilde{v}_1 = 0.454$ and $\tilde{v}_2 = 0.297$ THz).

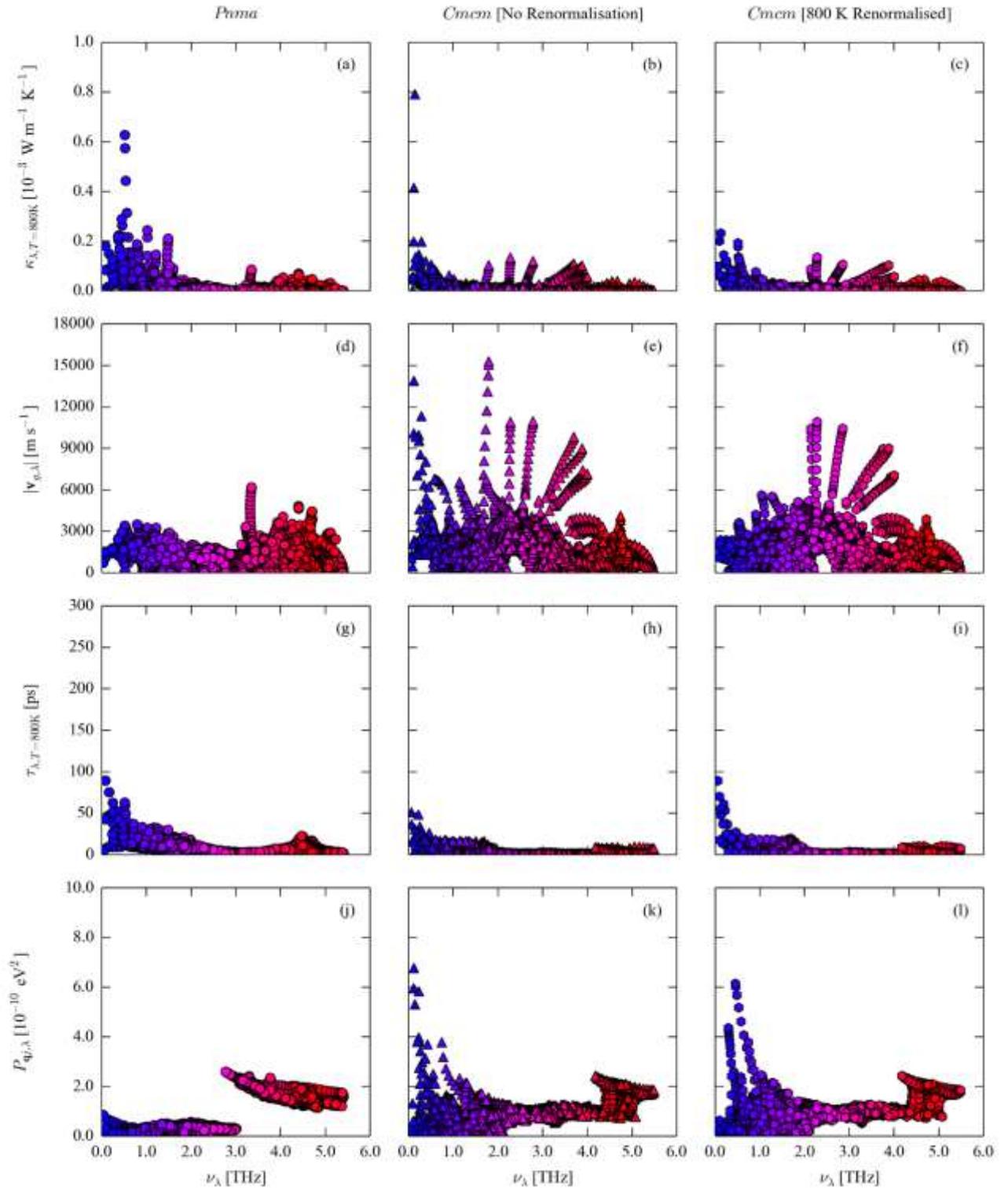

**Figure S14** Comparison of the isotropically-averaged modal contributions to the lattice thermal conductivity, $\kappa_\lambda$, of SnSe at 800 K (a-c) to the mode group velocities, $\left|\boldsymbol{v}_{g,\lambda}\right|$ (d-f), lifetimes, $\tau_\lambda$ (g-i), and averaged three-phonon interaction strengths, $P_{\mathbf{q}j,\lambda}$ (j-l; see Eq. S13). The left column shows data for the *Pnma* phase (left; a, d, g, j), while the other two show data for the *Cmcm* phase without renormalization of the two imaginary modes (center; b, e, h, k) and with the modes renormalized to the calculated 800 K frequencies (right; c, f, i, l; $\tilde{\nu}_1 = 0.454$ and $\tilde{\nu}_2 = 0.297$ THz).

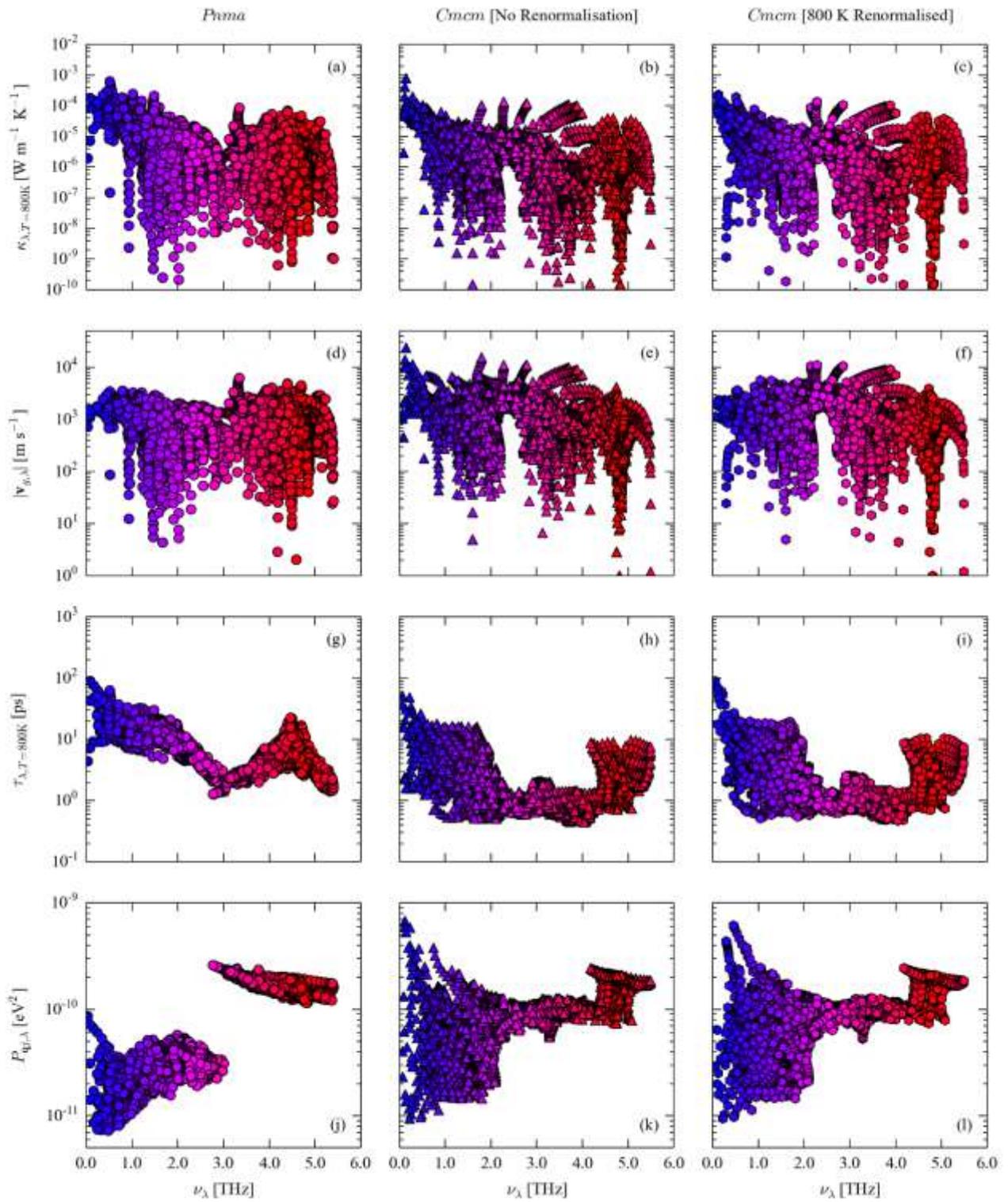

**Figure S15** Same data as in Fig. S14, plotted on a logarithmic scale.

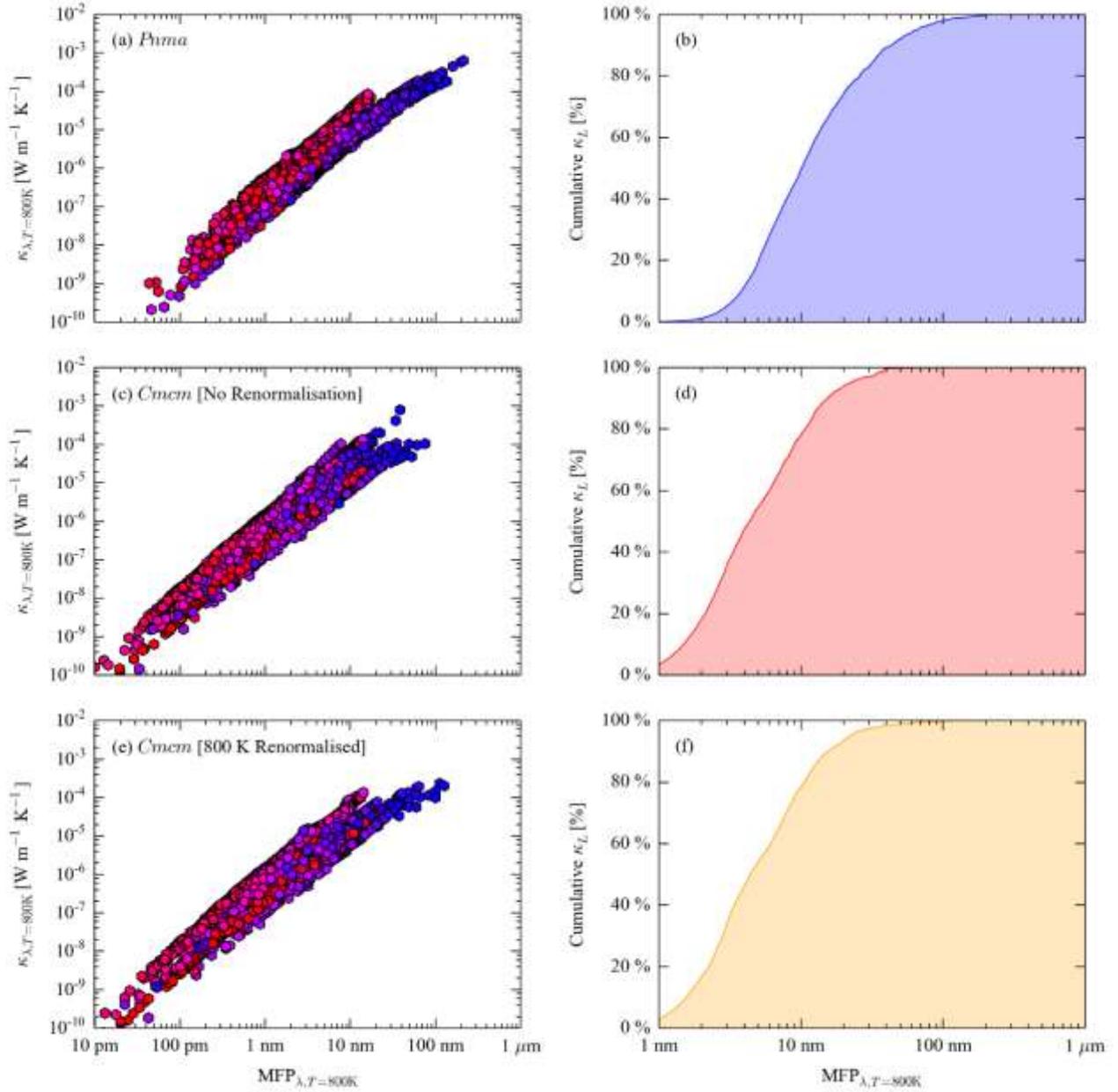

**Figure S16** Contributions of modes with different mean-free paths (MFPs) to the isotropically-averaged 800 K lattice thermal conductivity (a, c, e), together with the cumulative lattice thermal conductivity over MFPs between 1 nm and 1 μm (b, d, f), for *Pnma* (a, b) and *Cmcm* (c-f) SnSe. For the latter high-temperature phase, two sets of data are shown: the first is calculated without renormalization of the two soft modes (c, d), and the other with the modes renormalized to the calculated effective 800 K frequencies (e, f; $\tilde{v}_1 = 0.454$ and $\tilde{v}_2 = 0.297$ THz).